\newcommand{\myname}{Geoff Boeing, Yougeng Lu, and Clemens Pilgram}
\newcommand{\myemail}{boeing@usc.edu}
\newcommand{\myaffiliation}{University of Southern California}

\newcommand{\papertitle}{Local Inequities in the Relative Production of and Exposure to Vehicular Air Pollution in Los Angeles}
\newcommand{\papercitation}{Boeing, G., Y. Lu, and C. Pilgram. 2023. \papertitle. \textit{Urban Studies}, published online ahead of print. \href{https://doi.org/10.1177/00420980221145403}{doi:10.1177/00420980221145403}}
\newcommand{\paperkeywords}{air pollution, environmental justice, public health, race, spatial equity, transport planning}

\RequirePackage[l2tabu,orthodox]{nag}     
\documentclass[12pt,letterpaper]{article} 

\usepackage[T1]{fontenc}    
\usepackage[utf8]{inputenc} 
\usepackage{ebgaramond}     
\usepackage{tgheros}        

\usepackage[strict,autostyle]{csquotes} 
\usepackage[USenglish]{babel}           
\usepackage{microtype}                  

\usepackage{abstract}                   
\usepackage{amsmath}
\usepackage{array}
\usepackage{authblk}                    
\usepackage{booktabs}                   
\usepackage{caption}                    
\usepackage[final]{draftwatermark}      
\usepackage{endnotes}                   
\usepackage{geometry}                   
\usepackage{graphicx}                   
\usepackage{hyperref}                   
\usepackage{natbib}                     
\usepackage{pdflscape}                  
\usepackage{setspace}                   
\usepackage{titlesec}                   
\usepackage{url}                        
\usepackage{threeparttable}

\graphicspath{{./figures/}}

\titleformat{\section}{\normalfont\sffamily\large\bfseries\color{black}}{\thesection.}{0.3em}{}
\titleformat{\subsection}{\normalfont\sffamily\small\bfseries\color{black}}{\thesubsection.}{0.3em}{}

\hypersetup{
	pdfauthor={\myname},
	pdftitle={\papertitle},
	pdfsubject={\papertitle},
	pdfkeywords={\paperkeywords},
	pdffitwindow=true,         
	breaklinks=true,           
	colorlinks=false,          
	pdfborder={0 0 0}          
}

\begin{document}
	
\title{\papertitle\footnote{{Citation info: \papercitation}}}
\author[]{\myname\footnote{Corresponding author's email: \href{mailto:\myemail}{\myemail}}}
\affil[]{\myaffiliation}
\date{}

\maketitle

\begin{abstract}

Vehicular air pollution has created an ongoing air quality and public health crisis. Despite growing knowledge of racial injustice in exposure levels, less is known about the relationship between the production of and exposure to such pollution. This study assesses pollution burden by testing whether local populations' vehicular air pollution exposure is proportional to how much they drive. Through a Los Angeles, California case study we examine how this relates to race, ethnicity, and socioeconomic status---and how these relationships vary across the region. We find that, all else equal, tracts whose residents drive less are exposed to more air pollution, as are tracts with a less-White population. Commuters from majority-White tracts disproportionately drive through non-White tracts, compared to the inverse. Decades of racially-motivated freeway infrastructure planning and residential segregation shape today's disparities in who produces vehicular air pollution and who is exposed to it, but opportunities exist for urban planning and transport policy to mitigate this injustice.

\end{abstract}

\section{Introduction}

Twentieth century planners designed and constructed an enormous network of expressways to open up growing American metropolises to motorists. Vast swaths of established urban neighborhoods were bulldozed to clear new channels for suburban residents to drive to job centers. Yet some older neighborhoods survived relatively unscathed. For example, in Los Angeles, local residents organized to protest and eventually successfully cancel plans to extend State Route 2 through the affluent communities of Beverly Hills and Los Angeles's westside \citep{perez_angeles_2017}. In contrast, similar grassroots efforts failed in Los Angeles's eastside, where several major freeways carved up its less-affluent and less-White neighborhoods \citep{estrada_build_2005}. Race, wealth, and political power shaped the infrastructure planning that determines regional accessibility, travel behavior, and pollution exposure today.

Exposure to air pollution from cars, especially particulate matter of 2.5 microns or smaller (PM\textsubscript{2.5}), poses a direct risk to human health \citep{habre_contribution_2021, thompson_airborne_2018, sarzynski2012}. Much of the environmental justice literature around air pollution emphasizes community disparities in exposure, focusing on residential proximity to toxic release sites, refineries, road infrastructure, etc \citep{mikati_disparities_2018, schweitzer_environmental_2004, reichmuth_inequitable_2019, lievanos_racialized_2019}. These justice claims are straightforward when identifying a stationary emissions source, like a refinery: the refinery exposes nearby residents to harmful pollutants through a spatial diffusion process. Most empirical studies measure environmental risk through residential proximity to such stationary sources by assuming that proximity is a proxy for exposure \citep{doi:10.1111/0735-2166.00072, yuan_mega_2018}.

However, measuring exposure to emissions from mobile sources, such as cars, is less straightforward. Transport type and density along with local topography and meteorological conditions influence the degree, direction, and distance that air pollutants disperse \citep{LU2021110653, houston2004}. The political and planning decisions behind transport infrastructure placement impact environmental justice today through vehicular air pollution \citep{giles-corti_what_2022}. Importantly, the polluting vehicles may be driven by people with different sociodemographic profiles than the communities they traverse and pollute. Despite our growing knowledge of pollution exposure injustice \citep{mikati_disparities_2018, lievanos_racialized_2019, https://doi.org/10.1111/gean.12288}, we know less about the relationship between who generates vehicular air pollution and who is exposed to that pollution. The individuals exposed to the most pollution may simply be producing the most themselves. A stronger measure of environmental justice would ask: how much vehicular air pollution are you exposed to in relation to how much you drive?

This study asks this question in the context of race, class, and driving in Los Angeles County. Our data and methods are not unique to Los Angeles and can be applied elsewhere in the US. We focus on driving as roughly four in five of the county's commuters drive to work by themselves and another one in ten carpool \citep{county_of_los_angeles_open_data_commute_2019}. We model local exposure to vehicular air pollution as a function of vehicle kilometers traveled (VKT), accounting for the roles of race, income, and other covariates to better understand disparate infrastructure utilization and impacts. Using ordinary least squares (OLS) and geographically-weighted regression (GWR) we find that---all else equal---tracts that generate more vehicular travel tend to be exposed to less vehicular air pollution, and that non-White and poor communities are disproportionately burdened with excess pollution. Our GWR models in particular allow us to explain substantially more of this variation than the OLS models that are standard in the environmental justice literature. To illustrate how vehicular air pollution disparities unfold spatially, we simulate car commutes and develop a novel inequity index that measures the extent to which commutes through a tract are disproportionately made by drivers of a certain race or ethnicity. We find that commuters from majority-White tracts disproportionately drive through non-White tracts, compared to the inverse. Moving beyond the original placement of infrastructure, we conclude by discussing possible interventions such as VKT taxes, tolls, emissions standards, and electrification.

\section{Background}

Urban motor vehicle transport generates air pollution that can harm human health. This relationship is mediated by the historical placement and current utilization of transport infrastructure by motorists.

\subsection{Driving, Emissions, and Health}

Many studies have identified environmental externalities stemming from urban transport---particularly motor vehicles---including greenhouse gas emissions and air, water, and noise pollution. \citet{10.2307/20053837} estimated motor vehicles' environmental externalities and found that, compared to other environmental hazards, air pollution's costs are the greatest---particularly for human health.

Researchers have identified cars and trucks as the most significant sources of urban air pollution \citep{PMID:20560612, Rowangould2015ANA, doi:10.1080/10962247.2013.763867} and several recent studies have quantified the air pollution gradient near highways \citep{Brugge_Durant_Rioux_2007, doi:10.1177/0361198119825538, PMID:12269664}. The US Environmental Protection Agency (EPA) notes that transport accounts for a large share of PM\textsubscript{2.5} emissions \citep{us_epa_2021}. For example, in Southern California, vehicular emissions alone generate roughly a third of the region's PM\textsubscript{2.5} \citep{habre_contribution_2021, hasheminassab_long-term_2014}, and transport produces 37\% of California's GHG emissions---more than any other sector in the state \citep{CaliEmmissions}.

Air pollution correlates with an increased risk of lung cancer, asthma, bronchitis, and impaired cognitive development \citep{Sunyer2015, HUANG2019105167, jagai_county-level_2017}. A large body of literature has documented the deleterious effects of PM\textsubscript{2.5} exposure in particular \citep{feng_health_2016,thompson_airborne_2018,polichetti_effects_2009}. Several studies have demonstrated that children exposed to high-traffic roadways suffer from more respiratory ailments \citep{gauderman_childhood_2005, janssen_assessment_2001, oosterlee1996}. \citet{HUANG2019105167} found that higher daily PM\textsubscript{2.5} concentrations correspond to higher pediatric respiratory rates. PM\textsubscript{2.5} is associated with increased childhood asthma hospitalization \citep{https://doi.org/10.1111/j.1365-2222.2006.02555.x} and exposure to vehicular emissions can both cause and exacerbate asthma \citep{kunzli_breathless_2003}. Researchers have also found that children exposed to more vehicular air pollution demonstrate slower cognitive development \citep{Sunyer2015}. Further, air pollution exposure is associated with lung cancer incidence, especially among people who have never smoked \citep{jagai_county-level_2017}.

\subsection{Race and Class Disparities}

These health impacts pose an urgent problem that is unevenly distributed across populations. \citet{schweitzer_environmental_2004} argue that transport justice fundamentally concerns the distribution of transport costs and benefits. Environmental hazards disproportionately impact non-White and low-income individuals relative to Whiter and more affluent ones \citep{doi:10.1068/b32124, lievanos_racialized_2019, houston2004, Tessumeabf4491}. In particular, non-White and low-income communities are disproportionately burdened by transport externalities while simultaneously suffering from worse accessibility \citep{schweitzer_environmental_2004, rowangould_identifying_2016, yuan_mega_2018, poorfakhraei_evaluating_2017}. In tandem, this suggests that these communities do not benefit from transport infrastructure in proportion to the costs they bear.

Recent studies have used many different methods to measure racial inequity in air pollution exposure, including descriptive analysis \citep{brunt2017air, milojevic2017socioeconomic}, bivariate analysis \citep{rivas2017exposure, moreno2016assessing}, and multivariate analysis \citep{temam2017socioeconomic, padilla2014air, kim2021assessment}. Substantial heterogeneity exists across such studies in spatial scale (e.g., city, county, state), socioeconomic characteristics under consideration, and research design (e.g., cross-sectional or panel data). Yet one consistent finding in this literature has been that socially disadvantaged populations tend to be burdened with more air pollution exposure than other groups are.

The confluence of transport, health, and racial injustice is exemplified by Los Angeles, a poster child for automobile dependence, air pollution, racial diversity, and wealth inequality. According to the US Census Bureau, in 2019, 74\% of Los Angeles County residents were non-White. Latinos\endnote{Throughout, we use \enquote{Latino} as shorthand for the Census Bureau's designation of Hispanic or Latino of any race, \enquote{White} as shorthand for the Census Bureau's designation of non-Hispanic or Latino White alone, and \enquote{Black} as shorthand for the Census Bureau's designation of non-Hispanic or Latino Black or African-American alone.} (49\% of the total population) are the single largest group followed by Whites (26\%), Asians (15\%), and Blacks (9\%). The region's poverty rate is consistently higher than the national average. In Los Angeles, non-White and low-income populations are more likely to live in central neighborhoods for better accessibility and are exposed to twice the local traffic density as the rest of the region \citep{giuliano_travel_2003,houston2004}. Despite this exposure, they are less likely to own an automobile and more likely to commute by public transit \citep{policylink_equity_2017}. In other words, non-White and low-income populations live near more transport infrastructure yet benefit less from it.

Decades of racist planning decisions in Los Angeles have contributed to today's injustices in transport, health, and environmental quality. Twentieth century transport, housing, and land use planning accommodated White flight and industry at the expense of non-White communities \citep{estrada_build_2005, baum-snow_did_2007}. Planners bulldozed central neighborhoods to construct an expansive freeway system and create regional access for new peripheral suburbs and their predominately White residents \citep{pulido_rethinking_2000, boeing_off_2021}---often selecting freeway alignments through non-White neighborhoods for the sake of \enquote{slum clearance} or to protect business interests \citep{mohl_stop_2004, dimento_2009, perez_angeles_2017}. Freeway construction fragmented and displaced local residents and subjected surrounding communities to decades of subsequent air pollution \citep{mohl_stop_2004}. While some freeways were planned through Whiter and more affluent neighborhoods, they were often cancelled or rerouted due to these neighborhoods' political clout---such as with the aforementioned State Route 2 through Beverly Hills \citep{perez_angeles_2017}.

\section{Methods}

The literature suggests that some groups disproportionately benefit from automobility while other groups disproportionately bear its external costs. The present study advances this literature through a Los Angeles case study investigating the relative production of and exposure to vehicular air pollution, using neighborhood-scale aggregations. Are different communities exposed to vehicular air pollution at a level proportional to how much they drive? If not, what is the relationship between race and this disparity, all else equal? Such knowledge can help guide equitable planning efforts, target restorative policy interventions, and set specific environmental justice goals.

\begin{figure}[tb]
    \centering
    \includegraphics[width=\textwidth]{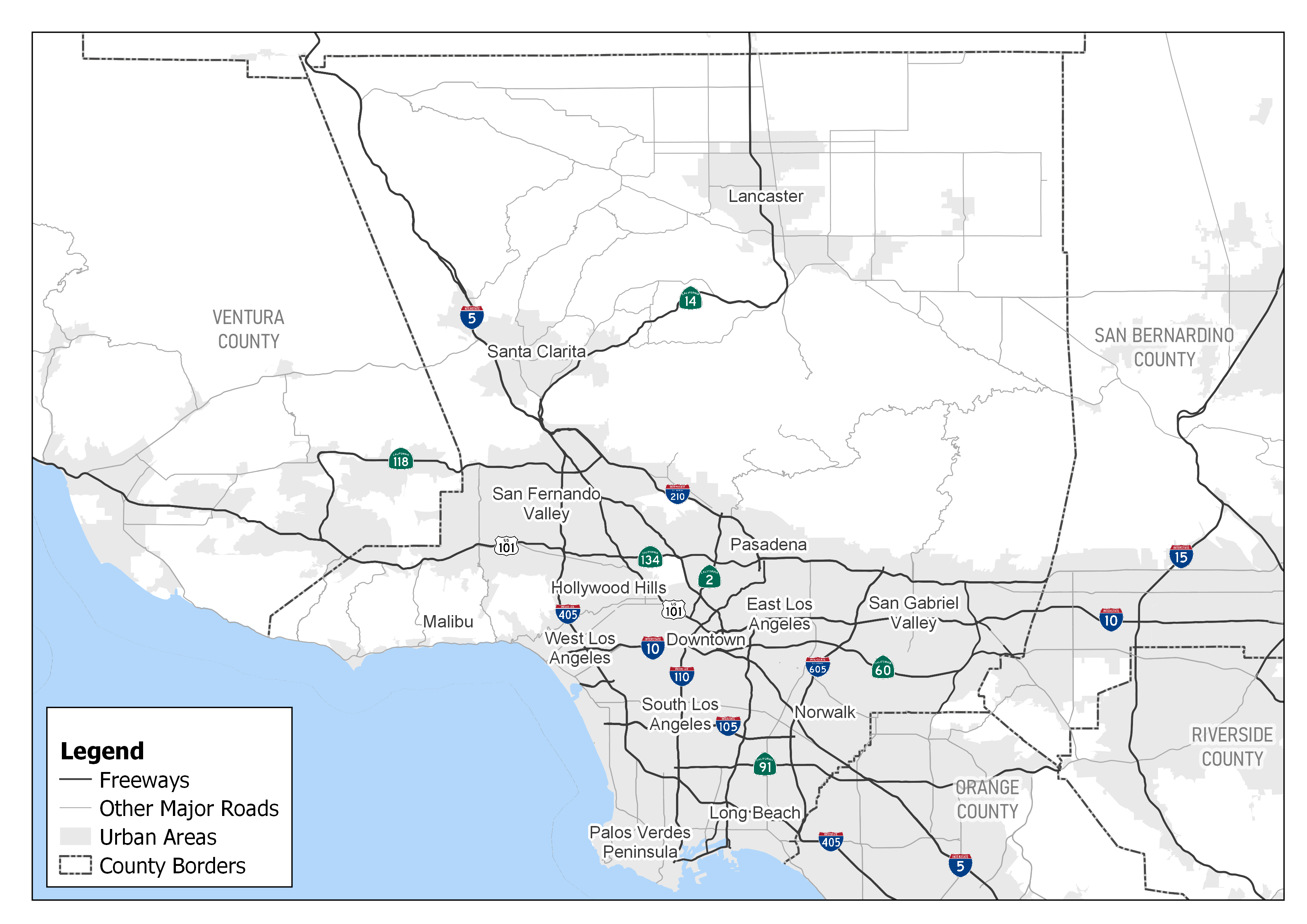}
    \caption{Los Angeles County's vicinity and highway infrastructure.}
    \label{fig:reference_map}
\end{figure}

\subsection{Data}

This study focuses on Los Angeles County (Figure~\ref{fig:reference_map}). Its sprawling structure, residential segregation, and transport infrastructure exemplify the spatial legacy of twentieth century American planning: today roughly 90\% of its commuters drive or carpool \citep{county_of_los_angeles_open_data_commute_2019}. We collect secondary data on air pollution, passenger vehicle travel, freight truck traffic, demographics, and street network design. We follow the literature by using census tract-level aggregations \citep[e.g.,][]{TAYARANI2020108999, LU2021111549, doi:10.1021/acs.est.6b02385}, as fully disaggregate pollution exposure data exist only for sample populations and are generally unavailable to the public. Tracts offer a useful unit of analysis as they roughly represent neighborhoods and follow real-world physical and social boundaries. Table~\ref{tab:variable_descriptions} describes our input data.

We use PM\textsubscript{2.5} concentrations from the Union of Concerned Scientists (UCS) as the primary air pollution exposure indicator. This data set combines emissions from on-road sources from the US EPA Emissions Inventory with the InMAP air pollution generation and transport model to estimate annual tract-level concentrations: \citet{reichmuth_inequitable_2019} and \citet{tessum_inmap_2017} detail this model and these data. This allows us to analyze local exposure to vehicular PM\textsubscript{2.5}. As a robustness check, we separately use CO\textsubscript{2} emissions from on-road sources from the Database of Road Transport Emissions (DARTE), in place of PM\textsubscript{2.5} concentrations \citep{https://doi.org/10.3334/ornldaac/1735}. While the PM\textsubscript{2.5} data measure concentration as the \enquote{stock} of air pollution, the CO\textsubscript{2} data measure tailpipe emissions as the \enquote{flow} of vehicular CO\textsubscript{2} into the air at a 1-kilometer resolution. Examining this alternative \enquote{flow} indicator checks the robustness of our primary \enquote{stock} indicator's results and provides insights into exposure to accumulated emissions by incorporating information on how pollutants disperse.

\begin{table}[tbp]
	\caption{Descriptions, units, and sources of tract-level variables. Source abbreviations are defined in the main text.}
	\label{tab:variable_descriptions}
	\footnotesize{}
	\begin{tabular}{m{0.25\linewidth} m{0.30\linewidth} m{0.20\linewidth} m{0.15\linewidth}}
		\toprule
		Variable                 & Description                                                       & Units                                         & Source            \\ \midrule
		PM\textsubscript{2.5}    & annual average on-road PM\textsubscript{2.5} concentration (2014) & $\mu$g/m\textsuperscript{3}                      & UCS               \\ \midrule
		CO$_{2}$                 & annual average road transport CO$_{2}$ emissions (2014)           & metric tons                                   & DARTE             \\ \midrule
		VKT                      & average daily vehicle travel generated by median household (2017) & kilometers                                    & LATCH             \\ \midrule
		Truck traffic volume     & daily distance traveled through tract by commercial trucks (2020) & kilometers                                    & SCAG              \\ \midrule
		Distance to highway      & distance from tract centroid to nearest highway (2019)            & kilometers                                    & Census TIGER/Line \\ \midrule
		Intersection density     & number of intersections per square kilometer (2020)               & intersections/km\textsuperscript{2}           & OpenStreetMap     \\ \midrule
		Grade Mean               & average street segment incline (2020)                             & n/a                                           & OpenStreetMap     \\ \midrule
		Proportion White         & proportion of White residents (2019)                              & n/a                                           & ACS               \\ \midrule
		Median household income  & annual median household income (2019)                             & 10,000s of inflation-adjusted 2018 USD        & ACS               \\ \midrule
		Proportion single-family & proportion of housing units that are single-unit detached (2019)  & n/a                                           & ACS               \\ \midrule
		Median rooms per home    & median rooms per housing unit (2019)                              & rooms                                         & ACS               \\ \midrule
		Mean household size      & average number of residents per housing unit (2019)               & residents                                     & ACS               \\ \midrule
		Population density       & residents per square kilometer (2019)                             & 1000s of residents/km\textsuperscript{2}      & ACS               \\ \midrule
		Median home value        & median value of owner-occupied housing units (2019)                      & 10,000s of inflation-adjusted 2018 USD & ACS               \\ \bottomrule
	\end{tabular}
\end{table}

We use tract-level resident VKT generation from the US Bureau of Transportation Statistics' Local Area Transportation Characteristics for Households (LATCH) to proxy vehicular air pollution production. This data set combines National Household Travel Survey responses with demographic data from the American Community Survey (ACS) to model household travel behavior and provide tract-level aggregations: the \citet{us_bureau_of_transportation_statistics_2017_2018} provides full methodological and validation details. Some temporal mismatch exists between the UCS PM\textsubscript{2.5} data's 2014 vintage and the LATCH VKT data's 2017 vintage. However, individual travel behavior and urban form change little over such a time span \citep{RAMEZANI202128}.

We simulate commuting trips (as detailed in Section 3.3) to determine driving routes, by race, through different tracts using the US Census Bureau's Longitudinal Employer-Household Dynamics (LEHD) Origin-Destination Employment Statistics (LODES). LODES is an administrative enumeration that includes private-sector employee home and work census blocks, offering a rough approximation, with some bias, of commute origins and destinations \citep{boeing_estimating_2018}. We aggregate these origins and destinations to the tract-level to match the rest of the input data.

We use modeled truck traffic volumes from the Southern California Association of Governments (SCAG) to control for freight truck traffic \citep{scag_heavy_2010}. We measure tract-level intersection density and street grade using OpenStreetMap data and the OSMnx software to control for differences in local street network structure \citep{boeing_osmnx_2017}. We use intersection density as a measure of street density and street grade as a measure of hilliness, which affects engine performance. We also collect tract-level race and income data from the 2018 5-year ACS, alongside a set of additional control variables summarized in Table~\ref{tab:variable_descriptions}. Some tracts lack observations across some of these variables. For consistent analysis across multiple specifications, we retain only those tracts with observations across all variables in Table~\ref{tab:variable_descriptions} ($n$=2238). Only 4\% of the county's tracts are thus excluded, but we note that they contain more multifamily housing, non-White residents, and lower income households than the retained tracts.

\subsection{Regression Analysis}

We estimate four regression models, two via OLS and two via GWR. Using OLS, we estimate the first two models as specified in Equation~\ref{eq:ols_model_log}:

\begin{equation}
    \label{eq:ols_model_log}
    \log\phi = \beta_{0} + \beta_{1} \log\tau + \beta X + \epsilon
\end{equation}

where $\log\phi$ is the response (log PM\textsubscript{2.5} concentration), $\beta_{0}$ is the intercept, $\log\tau$ is the log VKT generated by residents and $\beta_{1}$ is its coefficient to be estimated, $X$ is a matrix of $n$ observations on $k$ predictors and $\beta$ is its vector of coefficients to be estimated, and $\epsilon$ is the error term.

In Model 1, $X$ includes a limited set of controls including the White proportion of the population, median household income, truck traffic volume, and distance to the nearest highway. We define \enquote{highway} as interstate highways, US routes, and state routes, and measure the Euclidean distance between tract centroids and the nearest highway. In Model 2, $X$ includes a complete set of controls including all those from Model 1 as well as intersection density, mean street grade, proportion of single family homes, median number of rooms per home, mean household size, population density, and median home value. We log-transform predictors as needed for a linear relationship (as noted in Table~\ref{tab:regression_results}) and estimate all models with robust standard errors due to heteroskedasticity.

OLS estimates a global model with stationary parameters across the study region. However, such models cannot unpack spatially-varying statistical relationships. The literature suggests that air pollution exhibits spatial heterogeneity, but sophisticated models to investigate this remain rare in the literature. Given our interest in unpacking the potentially heterogeneous relationship between local PM\textsubscript{2.5} concentration and VKT generated, we estimate GWR models of the general form specified in Equation~\ref{eq:gwr_model} to observe any such local variation in coefficients and goodness-of-fit:

\begin{equation}
    \label{eq:gwr_model}
    \log\phi_{j} = \beta_{0j} + \beta_{1j} \log\tau_{j} + \beta_{j} X_{j} + \epsilon_{j}
\end{equation}

where $\log\phi_{j}$ is the response (log PM\textsubscript{2.5} concentration in each tract $j$), $\beta_{0j}$ is the intercept, $\beta_{1j}$ is a local coefficient to be estimated, $\log\tau_{j}$ is the log VKT generated by residents of $j$, $\beta_{j}$ is a vector of local coefficients to be estimated, $X_{j}$ is a matrix of observations in the local neighborhood of $j$, and $\epsilon_{j}$ is the local error term. In Model 3, $X_{j}$ includes the limited set of controls from Model 1. In Model 4, $X_{j}$ includes the complete set of controls from Model 2.

GWR estimates separate models for the local neighborhood of each tract in the study region. For each such local regression, observations are weighted by a Gaussian distance-decay function centered on $j$.  We use a spatially adaptive kernel that adjusts for the density of data at each regression location, due to the variability of census tract sizes across Los Angeles County. This determines a fixed number of nearest neighbors to adjust the bandwidth distance accordingly: tracts in dense areas have a narrower bandwidth and tracts in sparse areas have a wider one. We determine the optimal number of nearest neighbors according to the Akaike Information Criterion with small-sample correction (AICc) through iterative optimization. This results in 53 and 73 nearest neighbors for each local estimation of Models 3 and 4, respectively.

\subsection{Commute Simulation}

To examine demographic differences in driving patterns, we use block-level home and employment locations from LODES as a proxy to simulate commuters' routes to work. Although this is a microsimulation of commutes, we are interested in its aggregate outcomes rather than any specific individual commute. We solve routes by minimizing free-flow travel time, via Dijkstra's algorithm and OSMnx \citep{boeing_osmnx_2017}, from home block centroids to work block centroids. We identify, at the home tract level, all the other tracts through which its source trips pass. We assign each trip as White or non-White probabilistically given the home tract's White population proportion, and adjust the trip counts by the share of commuters who drive to work from each tract. Then we sum the total simulated kilometers driven through each tract, by White versus non-White commuters.

Finally, to investigate trends in the racial disparity of commutes through each tract, we developed the inequity index defined in Equation \ref{eq:inequity_index}:

\begin{equation}
    \label{eq:inequity_index}
    I_{jg} = \frac{D_{jg}}{D_{j}}-\frac{C_{jg}}{C_{j}}
\end{equation}

where the index $I_{jg}$ quantifies the extent to which commuter vehicular distances traveled through tract $j$ are disproportionately made by subgroup $g$ (either White or non-White). $D_{jg}$ is the distance traveled through $j$ by $g$, and $D_{j}$ is the total distance traveled through $j$ by all subgroups. $C_{jg}$ is the number of $g$ commuters living in $j$, and $C_{j}$ is the total number of commuters living in $j$. Positive and negative values of $I_{jg}$ indicate, respectively, disproportionately high and low distances traveled by subgroup $g$ through tract $j$.

\section{Results}

\subsection{OLS Results}

Table~\ref{tab:regression_results} reports the OLS results for Models 1 and 2, revealing a significant negative relationship between a tract's vehicular PM\textsubscript{2.5} concentration and its residents' VKT generation. Controlling for race, income, truck traffic, and highway proximity in Model 1, a 1\% increase in VKT generation is associated with a 1.24\% decrease in local PM\textsubscript{2.5} exposure. The full set of controls in Model 2 moderates the magnitude, but a 1\% increase in VKT is still associated with a 0.62\% decrease in local PM\textsubscript{2.5} exposure. All else equal, tracts that generate \textit{more} vehicular travel tend to be exposed to \textit{less} vehicular air pollution---an important paradox we unpack in the discussion below.

Both Models 1 and 2 reveal a significant negative relationship between a tract's vehicular PM\textsubscript{2.5} concentration and the White proportion of the population. Even when controlling for income, home value, truck traffic, and other covariates, Whiter tracts tend to be exposed to less vehicular air pollution. In other words, non-White communities, whether high-income or low-income, are exposed to more PM\textsubscript{2.5} than otherwise-similar White communities. These models demonstrate that vehicular air pollution burdens distribute inequitably with regard to race across Los Angeles County as a whole.

\begin{landscape}
    \begin{table}[htbp]
        \centering
        \small
        \caption{Regression model parameter estimates for Model 1 (basic OLS), Model 2 (OLS with full controls), Model 3 (basic GWR), and Model 4 (GWR with full controls). Standard error is shown in parentheses. Significance is denoted as * $p<0.05$.}
        \label{tab:regression_results}
        \footnotesize
        \begin{tabular}{lrrrrrrrrrrrr}
            \toprule
            & \multicolumn{1}{c}{(1)} & \multicolumn{1}{c}{(2)} & \multicolumn{5}{c}{(3)} & \multicolumn{5}{c}{(4)} \\
            \cmidrule(lr){2-2}\cmidrule(lr){3-3}\cmidrule(lr){4-8}\cmidrule(lr){9-13} &  \multicolumn{1}{c}{estimate}  &   \multicolumn{1}{c}{estimate}    & \multicolumn{1}{c}{mean} & \multicolumn{1}{c}{min} & \multicolumn{1}{c}{max} & \multicolumn{1}{c}{$t$<-1.96} & \multicolumn{1}{c}{$t$>1.96} & \multicolumn{1}{c}{mean} & \multicolumn{1}{c}{min} & \multicolumn{1}{c}{max} & \multicolumn{1}{c}{$t$<-1.96} & \multicolumn{1}{c}{$t$>1.96} \\
            \midrule
            Intercept                 &  6.13* &  1.31* &  2.02 & -1.58 & 11.00 &  2.32\% & 70.60\% &  1.37 & -1.98 & 4.84 &  2.32\% & 57.42\% \\
                                      & (0.29) & (0.38) &       &       &       &         &         &       &       &      &  \\
            VKT (log)                 & -1.24* & -0.62* & -0.23 & -2.79 &  0.68 & 38.52\% &  8.40\% & -0.18 & -1.46 & 0.54 & 33.91\% &  6.52\% \\
                                      & (0.08) & (0.10) &       &       &       &         &         &       &       &      &  \\
            Proportion White          & -0.61* & -0.63* & -0.14 & -1.36 &  0.76 & 32.17\% & 20.38\% & -0.03 & -0.95 & 1.22 & 27.84\% & 24.98\% \\
                                      & (0.06) & (0.07) &       &       &       &         &         &       &       &      &  \\
            Median household income   &  0.04* &  0.01~~& -0.01 & -0.09 &  0.25 & 38.87\% & 11.66\% &  0.00 & -0.06 & 0.10 &  8.76\% &  7.95\% \\
                                      & (0.01) & (0.01) &       &       &       &         &         &       &       &      &  \\
            Truck traffic volume (log)& -0.01* &  0.01~~&  0.00 & -0.05 &  0.05 &  6.88\% & 13.94\% &  0.01 & -0.05 & 0.07 &  7.86\% & 21.94\% \\
                                      & (0.01) & (0.00) &       &       &       &         &         &       &       &      &  \\
            Distance to highway       & -0.12* & -0.09* & -0.05 & -0.15 &  0.02 & 81.14\% &  0.09\% & -0.04 & -0.12 & 0.02 & 76.68\% &  0.09\% \\
                                      & (0.01) & (0.01) &       &       &       &         &         &       &       &      &  \\
            Intersection density      &        & -0.00~~&       &       &       &         &         &  0.00 & -0.01 & 0.00 & 11.21\% & 11.17\% \\
                                      &        & (0.00) &       &       &       &         &         &       &       &      &  \\
            Grade mean (log)          &        & -0.06* &       &       &       &         &         & -0.04 & -0.39 & 0.14 & 37.85\% & 19.71\% \\
                                      &        & (0.01) &       &       &       &         &         &       &       &      &  \\
            Prop single-family        &        &  0.44* &       &       &       &         &         &  0.07 & -0.96 & 0.83 &  3.66\% & 19.84\% \\
                                      &        & (0.07) &       &       &       &         &         &       &       &      &  \\
            Median rooms per home     &        & -0.22* &       &       &       &         &         & -0.05 & -0.44 & 0.12 & 43.21\% &  7.19\% \\
                                      &        & (0.02) &       &       &       &         &         &       &       &      &  \\
            Mean household size       &        &  0.09* &       &       &       &         &         &  0.04 & -0.20 & 0.33 & 10.90\% & 38.20\% \\
                                      &        & (0.02) &       &       &       &         &         &       &       &      &  \\
            Population density (log)  &        &  0.07* &       &       &       &         &         &  0.02 & -0.10 & 0.19 &  1.30\% & 19.57\% \\
                                      &        & (0.02) &       &       &       &         &         &       &       &      &  \\
            Median home value (log)   &        &  0.66* &       &       &       &         &         &  0.05 & -0.41 & 1.19 & 15.19\% & 23.82\% \\
                                      &        & (0.03) &       &       &       &         &         &       &       &      &  \\ \midrule
            $R^2$                     &   0.38 &   0.54 &  0.67 &  0.00 &  0.94 &         &         &  0.74 &  0.00 & 0.97 &     &         \\ \bottomrule
        \end{tabular}
    \end{table}
\end{landscape}

Re-estimating these models using (log) on-road CO\textsubscript{2} emissions as the response tells a similar story. A 1\% increase in VKT generation is associated with a 0.52\% decrease in local CO\textsubscript{2} exposure, and 0.38\% when including the full set of controls. A significant negative relationship is similarly found between the tracts' CO\textsubscript{2} exposure and their White population proportion.

\subsection{GWR Results}

Table~\ref{tab:regression_results} also reports the GWR results for Models 3 and 4 by summarizing the distributions of their coefficients and goodness-of-fit measures. The GWR models perform much better than the OLS models due to spatial heterogeneity in the relationships studied. While Model 1 explains only 38\% of the response's variance, Model 3 (with the same predictors) explains 67\% of its local variance on average. Similarly,  Model 2 explains 54\% of the response's variance, but Model 4 explains 74\% locally on average.

These GWR models unpack the spatial variation of individual predictors' relationships across the region. For example, in Model 1, the White proportion of the population is (uniformly) significantly and negatively associated with vehicular PM\textsubscript{2.5} exposure. However, Model 3 reveals a more nuanced and spatially varying relationship. Its coefficients for the White proportion of the population range from -1.36 to 0.76 with a mean value of -0.14 (see Table~\ref{tab:regression_results}). The relationship between these variables is not stationary across our study region, though it tends to be negative on average---similar to what Models 1 and 2 suggest globally. One potential explanation for this spatial variation could be sorting: people choose to live in different places for different reasons, including heterogeneous distaste for exposure to emissions and differences in how they value accessibility to other locations. These variations in coefficient significance and direction highlight the importance of assessing these relationships locally as well as globally.

\begin{figure}[tb]
\centering
\includegraphics[width=\textwidth]{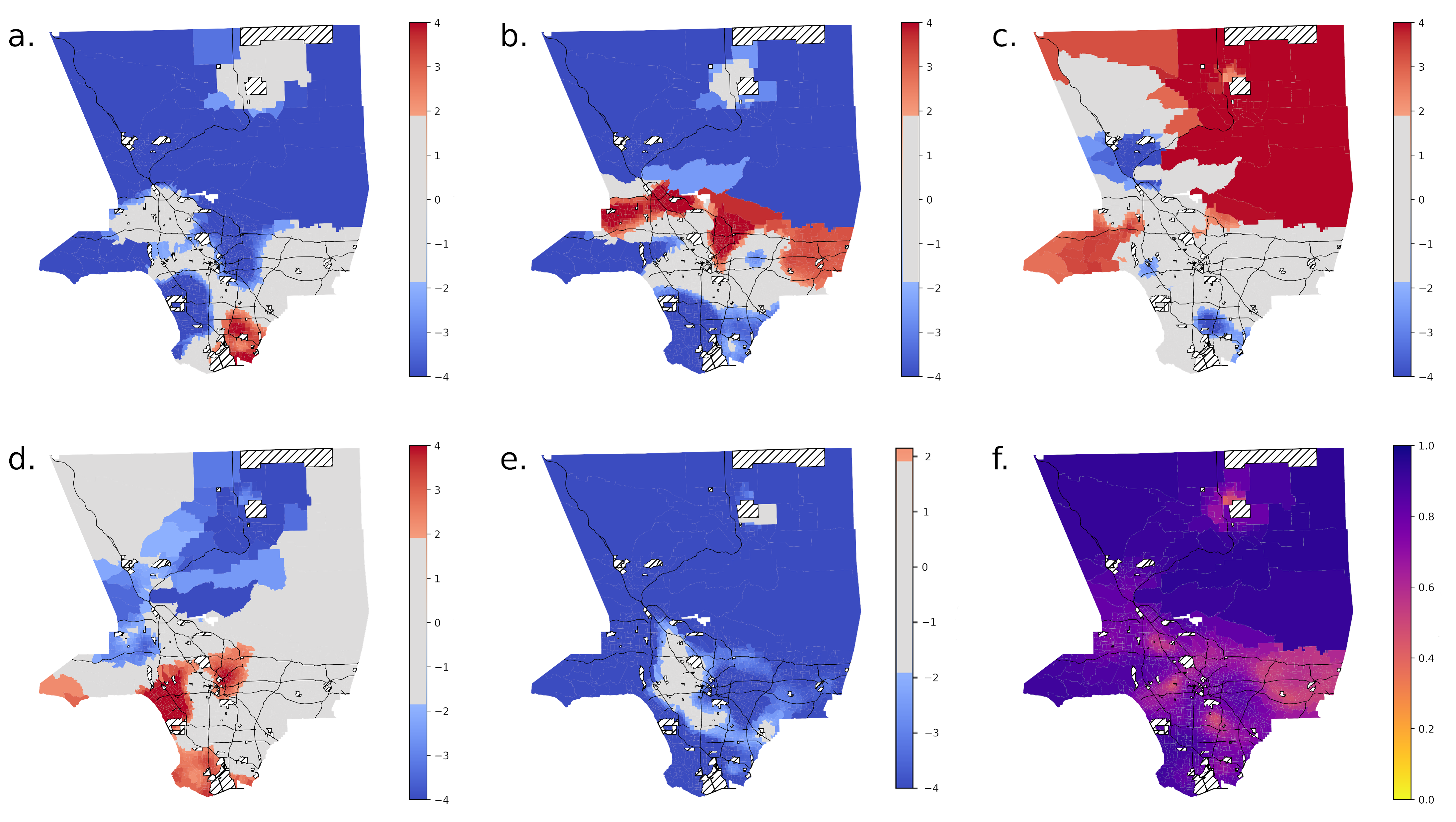}
\caption{Spatial distribution of Model 4's GWR local $t$-statistics for a) log VKT, b) proportion White, c) median household income, d) log truck traffic volume, e) distance to highway, and f) local $R^2$.}
\label{fig:gwr_model_4_distributions}
\end{figure}

Figure~\ref{fig:gwr_model_4_distributions} depicts Model 4's spatial distributions of local $t$-statistics and $R^2$ values across the study region. The combined statistical relationship of the predictors with the response varies spatially and this figure illustrates where the GWR models offer a better fit than the OLS models could. Overall, the GWR $R^2$ values improve on the OLS values in most tracts.

The consistency of these model results across specifications lends confidence in their estimates. In Model 3, 39\% of tracts show a significant, negative relationship between resident VKT generation and vehicular PM\textsubscript{2.5} exposure, while only 8\% of tracts show a significant, positive relationship. With the full set of controls in Model 4, 34\% of tracts show a significant, negative relationship between resident VKT generation and vehicular PM\textsubscript{2.5} exposure, while only 7\% of tracts show a significant, positive relationship. Figure~\ref{fig:gwr_model_4_distributions}a illustrates where these relationships tend to be negative, including the low-income Latino eastside of Los Angeles and the high-income White communities along the oceanfront. Such areas demonstrate the greatest injustice as those who drive more are exposed to less pollution, and vice versa.

\begin{figure}[tb]
\centering
\includegraphics[width=\textwidth]{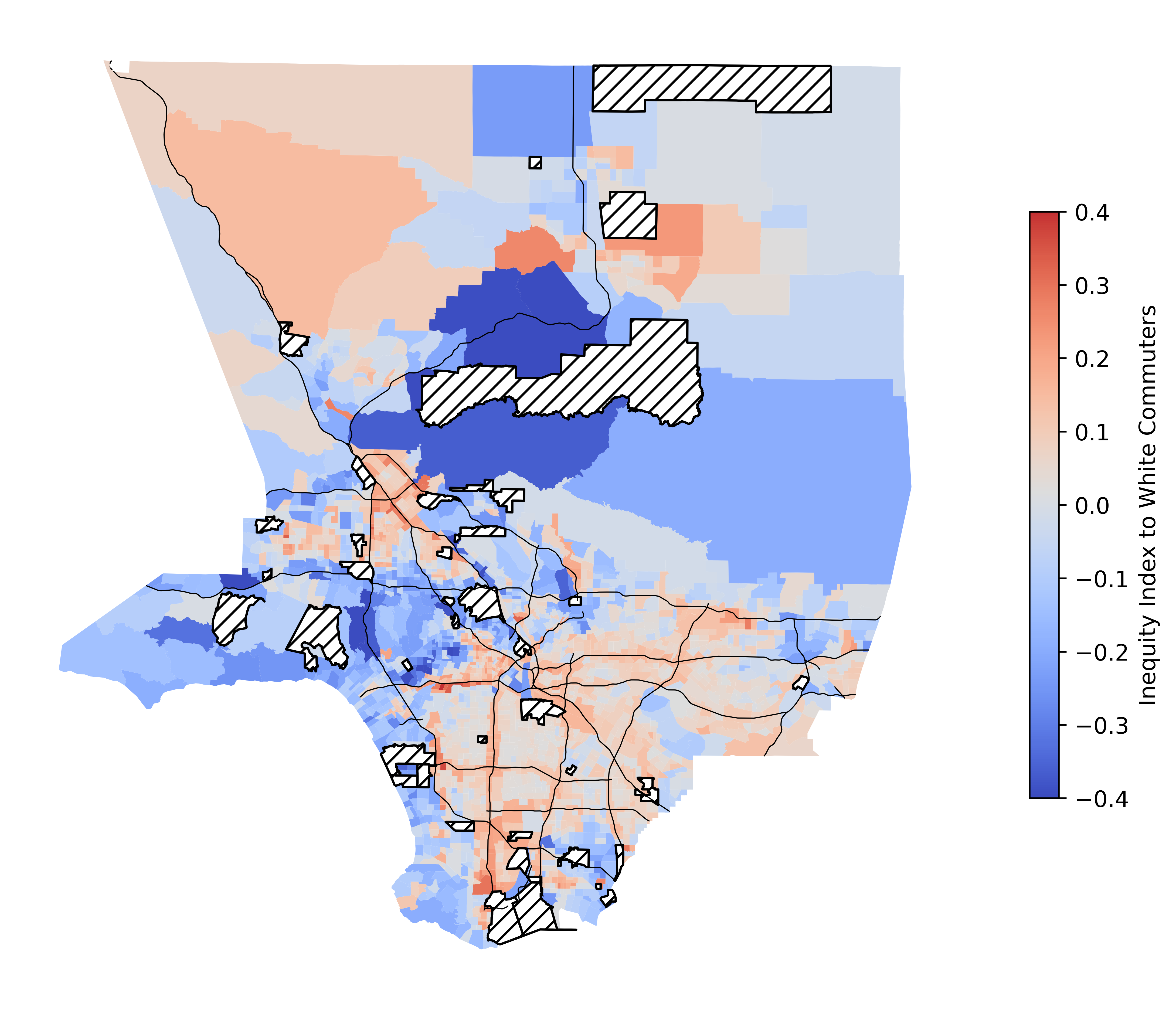}
\caption{Tract-level inequity index: positive values indicate that the share of White commuters traversing the tract exceeds the share of White residents living in the tract. Negative values indicate the opposite.}
\label{fig:inequity_index}
\end{figure}

\subsection{Commute Simulation}

The commute simulation results reveal unequal patterns in driving between White and non-White commuters which help explain the aforementioned exposure disparity. Figure~\ref{fig:inequity_index} depicts the inequity index's spatial distribution. Positive values indicate that commuters residing in a tract are less White than the commuters driving through the tract, while negative values indicate the opposite. It does not show \textit{where} driving occurs: most driving is constrained to highway corridors (and thus non-White areas). Accordingly, more trips traverse the figure's red tracts than blue tracts.

Countywide, the inequity index's population-weighted mean is 0.0153, revealing that most tracts experienced disproportionately more travel by White commuters relative to the local racial composition. This disparity is more than twice as large in tracts with highways (0.0307) than in those without (0.0147). Figure~\ref{fig:inequity_index} illustrates the roles of residential segregation and highway placement in today's driving patterns. Major highways stand out in particular. Tracts adjacent to Interstate 10---an east-west backbone---have a population-weighted mean of 0.0957, while those adjacent to Interstates 110 and 105---the largest freeways through predominately Black and Latino South Los Angeles---have a value of 0.1288. Thus, commuters driving through these tracts are respectively 10 and 13 percentage points Whiter than the local resident population---particularly notable when considering that the countywide population is less than one-third White. Yet there are also examples of major highways on which commuters resemble or are less White than local residents. Tracts adjacent to Interstate 405 and US Route 101 have inequity index values of -0.0012 and -0.0999, respectively. Both highways traverse mountain passes through the majority-White, affluent Hollywood Hills and Santa Monica Mountains where alternative alignments were impossible or cost-prohibitive.

\section{Discussion}

The research literature has explored race and class disparities in local vehicular air pollution exposure without accounting for local residents' own contributions to their exposure. This study finds that---all else equal---tracts whose residents drive less are exposed to more vehicular air pollution. Furthermore, tracts with a larger non-White population proportion---whether high- or low-income---experience more air pollution than do Whiter but otherwise similar tracts. Comparable trends hold in the GWR models on average as well. This reveals an injustice in pollution burden with a distinct racial dimension and these results are robust across multiple specifications and both global and local estimations. Overall, our findings reveal a systematic environmental injustice: road infrastructure benefits Whiter communities while exacting a cost on less-White communities.

This study opens the door for future research. While the simulation analysis provides evidence that unequal driving patterns could generate some of the observed disparity, it does not rule out the influence of other mechanisms such as tract-level differences in wind patterns or fuel efficiency. Future research should continue investigating these relationships. Nevertheless, our models consistently reveal the same broad story: tracts with a greater share of White residents and tracts with residents who drive more are exposed to less vehicular air pollution. One natural way to interpret this relationship is that people from majority-White neighborhoods do most of their \enquote{excess} driving through non-White neighborhoods and are thus exposed to less pollution at home despite producing more of it. However, there are other potential mechanisms that could explain the relationship. For example, majority-White neighborhoods could be located in areas where prevailing winds push pollution away and toward non-White neighborhoods, or majority-White neighborhoods' residents could drive more fuel-efficient cars so that even though they drive more, they produce and are exposed to less pollution at home. However, the history of highway construction through non-White neighborhoods suggests that differences in where travel occurs at least partially drive these results.

Further, the commute simulation provides evidence that disparities in vehicular air pollution exposure at home result from where people drive. On average, White commuters traverse tracts that are far more non-White than the tracts where most White commuters live. This disparity does not exist in the opposite direction: on average, non-White commuters do not travel through tracts that are substantially Whiter than their home tracts. This also illustrates the role that highways play in mediating this disparity. When White commuters traverse non-White tracts, they do so predominantly through tracts that contain highways. In other words, White commuters receive the benefits of driving on a highway, but because those highways are predominantly in non-White neighborhoods, other racial groups bear external costs of that driving.

This disparity is less common in the opposite direction. On average, non-White commuters do not travel through tracts that are substantially Whiter than their home tracts---and even where they occasionally do, it is on freeway segments that follow alignments determined by topography. For example, by \textit{not} building a freeway through Beverly Hills, planners caused a substantial share of commuters who would have taken this route to instead drive through majority non-White tracts on Interstate 10 instead. However, a similar rerouting of Interstate 405 or US Route 101 would not be possible without much greater and costlier detours. Transport planning has played a central---but not solo---role in these inequities: land use regulation, zoning, and housing policy interact with transport plans to generate these outcomes.

Several policies can help mitigate this environmental injustice. First, policymakers could continue raising fuel efficiency standards for new cars and encouraging vehicular electrification. Both would reduce on-road emissions without necessarily reducing the amount of driving. However, they would not eliminate rubber tires' and brake dust's substantial contributions to PM\textsubscript{2.5} pollution. Second, policymakers could enact tolls or other forms of congestion taxes to reduce total driving or capture its externalities. Third, policymakers could discourage commuting altogether by incentivizing more people to work from home, such as through tax credits. The Covid-19 pandemic witnessed a surge in remote work. Continuing these work-from-home trends after the pandemic could reduce vehicular travel and air pollution, particularly by higher-income White collar workers with more flexible jobs. Finally, policymakers can address environmental injustice through the housing market. Permitting more residential construction in job-rich neighborhoods could reduce commute distances. Further, legalizing the construction of denser and more affordable housing in less-polluted, exclusive neighborhoods could reduce exposure disparities. Yet no single policy can eliminate longstanding systemic discrimination against low-income and non-White populations.

\section{Conclusion}

This paper extended quantitative research on transport-environmental justice through a spatial analysis of the production of and exposure to vehicular air pollution in Los Angeles. We assessed local exposure to vehicular air pollution adjusted by local production of VKT through four models and two estimation techniques. In particular, our GWR analysis allowed for local nuance to reveal spatial heterogeneity.

We found that tracts that generate more vehicular travel tend to be exposed to less vehicular air pollution, all else equal. Race significantly predicts pollution exposure, even when controlling for a full set of related variables. Our commute simulation's inequity index helps explain these findings by illustrating the role of freeways and residential sorting in commute patterns. On average, White commuters traverse tracts that are far more non-White than their home tracts, but non-White commuters do not travel through tracts that are substantially Whiter than their own. Dismantling decades of racially-motivated transport planning and segregation requires concerted effort by planners and policymakers to redress past harms and envision a more equitable future.

\section*{Acknowledgments}

This research was supported by a grant from the US Department of Transportation and the Pacific Southwest Region University Transportation Center. The authors wish to thank Peter Mannino, Nicholas Cerdera, and David Flores Moctezuma for research assistance.

\IfFileExists{\jobname.ent}{\theendnotes}{}

\setlength{\bibsep}{0.00cm plus 0.05cm} 
\bibliographystyle{apalike}
\bibliography{references}

\begin{thebibliography}{}

\bibitem[Bae et~al., 2007]{doi:10.1068/b32124}
Bae, C.-H.~C., Sandlin, G., Bassok, A., and Kim, S. (2007).
\newblock The exposure of disadvantaged populations in freeway air-pollution
  sheds: A case study of the {Seattle} and {Portland} regions.
\newblock {\em Environment and Planning B}, 34(1):154--170.

\bibitem[Baum-Snow, 2007]{baum-snow_did_2007}
Baum-Snow, N. (2007).
\newblock Did {Highways} {Cause} {Suburbanization}?
\newblock {\em The Quarterly Journal of Economics}, 122(2):775--805.

\bibitem[Boeing, 2017]{boeing_osmnx_2017}
Boeing, G. (2017).
\newblock {OSMnx}: New methods for acquiring, constructing, analyzing, and
  visualizing complex street networks.
\newblock {\em Computers, Environment and Urban Systems}, 65:126--139.

\bibitem[Boeing, 2018]{boeing_estimating_2018}
Boeing, G. (2018).
\newblock Estimating local daytime population density from census and payroll
  data.
\newblock {\em Regional Studies, Regional Science}, 5:179--182.

\bibitem[Boeing, 2021]{boeing_off_2021}
Boeing, G. (2021).
\newblock Off the {Grid}… and {Back} {Again}? {The} {Recent} {Evolution} of
  {American} {Street} {Network} {Planning} and {Design}.
\newblock {\em Journal of the American Planning Association}, 87(1):123--137.

\bibitem[Brugge et~al., 2007]{Brugge_Durant_Rioux_2007}
Brugge, D., Durant, J.~L., and Rioux, C. (2007).
\newblock Near-highway pollutants in motor vehicle exhaust: A review of
  epidemiologic evidence of cardiac and pulmonary health risks.
\newblock {\em Environmental Health}, 6(1):23.

\bibitem[Brunt et~al., 2017]{brunt2017air}
Brunt, H., Barnes, J., Jones, S., Longhurst, J., Scally, G., and Hayes, E.
  (2017).
\newblock Air pollution, deprivation and health: understanding relationships to
  add value to local air quality management policy and practice in {Wales},
  {UK}.
\newblock {\em Journal of Public Health}, 39(3):485--497.

\bibitem[{Bureau of Transportation Statistics},
  2018]{us_bureau_of_transportation_statistics_2017_2018}
{Bureau of Transportation Statistics} (2018).
\newblock 2017 {Local} {Area} {Transportation} {Characteristics} for
  {Households} {Methodology}.

\bibitem[{California Air Resources Board}, 2020]{CaliEmmissions}
{California Air Resources Board} (2020).
\newblock Current {California} {GHG} {Emission} {Inventory} {Data}.

\bibitem[{County of Los Angeles Open Data},
  2019]{county_of_los_angeles_open_data_commute_2019}
{County of Los Angeles Open Data} (2019).
\newblock Commute {Mode} {Share} in {LA} {County} (2005-2017).

\bibitem[Delucchi, 2000]{10.2307/20053837}
Delucchi, M.~A. (2000).
\newblock Environmental externalities of motor-vehicle use in the {US}.
\newblock {\em Journal of Transport Economics and Policy}, 34(2):135--168.

\bibitem[DiMento, 2009]{dimento_2009}
DiMento, J. F.~C. (2009).
\newblock Stent (or dagger?) in the heart of town: Urban freeways in
  {Syracuse}, 1944—1967.
\newblock {\em Journal of Planning History}, 8(2):133--161.

\bibitem[Estrada, 2005]{estrada_build_2005}
Estrada, G. (2005).
\newblock If you build it, they will move: The {Los Angeles} freeway system and
  the displacement of {Mexican East Los Angeles}, 1944-1972.
\newblock {\em Southern California Quarterly}, pages 287--315.

\bibitem[Feng et~al., 2016]{feng_health_2016}
Feng, S., Gao, D., Liao, F., Zhou, F., and Wang, X. (2016).
\newblock The health effects of ambient {PM2}.5 and potential mechanisms.
\newblock {\em Ecotoxicology and Environmental Safety}, 128:67--74.

\bibitem[Gately et~al., 2019]{https://doi.org/10.3334/ornldaac/1735}
Gately, C., Hutyra, L., and Wing, I. (2019).
\newblock {DARTE Annual On-road CO2 Emissions on a 1-km Grid, Conterminous USA
  V2, 1980-2017}.

\bibitem[Gauderman et~al., 2005]{gauderman_childhood_2005}
Gauderman, W.~J., Avol, E., Lurmann, F., Kuenzli, N., Gilliland, F., Peters,
  J., and McConnell, R. (2005).
\newblock Childhood {Asthma} and {Exposure} to {Traffic} and {Nitrogen}
  {Dioxide}:.
\newblock {\em Epidemiology}, 16(6):737--743.

\bibitem[Giles-Corti et~al., 2022]{giles-corti_what_2022}
Giles-Corti, B., Moudon, A.~V., Lowe, M., Cerin, E., Boeing, G., Frumkin, H.,
  Salvo, D., Foster, S., Kleeman, A., Bekessy, S., de~Sá, T.~H.,
  Nieuwenhuijsen, M., Higgs, C., Hinckson, E., Adlakha, D., Arundel, J., Liu,
  S., Oyeyemi, A.~L., Nitvimol, K., and Sallis, J.~F. (2022).
\newblock What next? {Expanding} our view of city planning and global health,
  and implementing and monitoring evidence-informed policy.
\newblock {\em The Lancet Global Health}, 10(6):e919--e926.

\bibitem[Giuliano, 2003]{giuliano_travel_2003}
Giuliano, G. (2003).
\newblock Travel, location and race/ethnicity.
\newblock {\em Transportation Research Part A}, 37:351--372.

\bibitem[Habre et~al., 2021]{habre_contribution_2021}
Habre, R., Girguis, M., Urman, R., Fruin, S., Lurmann, F., Shafer, M., Gorski,
  P., Franklin, M., McConnell, R., Avol, E., and Gilliland, F. (2021).
\newblock Contribution of tailpipe and non-tailpipe traffic sources to
  quasi-ultrafine, fine and coarse particulate matter in {Southern California}.
\newblock {\em Journal of the Air \& Waste Management Association},
  71(2):209--230.

\bibitem[Hasheminassab et~al., 2014]{hasheminassab_long-term_2014}
Hasheminassab, S., Daher, N., Ostro, B.~D., and Sioutas, C. (2014).
\newblock Long-term source apportionment of ambient fine particulate matter
  ({PM} 2.5 ) in the {Los} {Angeles} {Basin}: {A} focus on emissions reduction
  from vehicular sources.
\newblock {\em Environmental Pollution}, 193:54--64.

\bibitem[Houston et~al., 2004]{houston2004}
Houston, D., Wu, J., Ong, P., and Winer, A. (2004).
\newblock {Structural Disparities of Urban Traffic in Southern California:
  Implications for Vehicle-Related Air Pollution Exposure in Minority and
  High-Poverty Neighborhoods}.
\newblock {\em Journal of Urban Affairs}, 26:565--592.

\bibitem[Huang et~al., 2019]{HUANG2019105167}
Huang, M., Ivey, C., Hu, Y., Holmes, H.~A., and Strickland, M.~J. (2019).
\newblock {Source apportionment of primary and secondary PM2.5: Associations
  with pediatric respiratory disease emergency department visits in the U.S.
  State of Georgia}.
\newblock {\em Environment International}, 133:105167.

\bibitem[Jagai et~al., 2017]{jagai_county-level_2017}
Jagai, J.~S., Messer, L.~C., Rappazzo, K.~M., Gray, C.~L., Grabich, S.~C., and
  Lobdell, D.~T. (2017).
\newblock County-level cumulative environmental quality associated with cancer
  incidence: {Environment} and {Cancer} {Incidence}.
\newblock {\em Cancer}, 123(15):2901--2908.

\bibitem[Janssen et~al., 2001]{janssen_assessment_2001}
Janssen, N.~A., van Vliet, P.~H., Aarts, F., Harssema, H., and Brunekreef, B.
  (2001).
\newblock Assessment of exposure to traffic related air pollution of children
  attending schools near motorways.
\newblock {\em Atmospheric Environment}, 35(22):3875--3884.

\bibitem[Karner et~al., 2010]{PMID:20560612}
Karner, A.~A., Eisinger, D.~S., and Niemeier, D.~A. (2010).
\newblock Near-roadway air quality: Synthesizing the findings from real-world
  data.
\newblock {\em Environmental Science \& Technology}, 44(14):5334—5344.

\bibitem[Kim and Kwan, 2021]{kim2021assessment}
Kim, J. and Kwan, M.-P. (2021).
\newblock Assessment of sociodemographic disparities in environmental exposure
  might be erroneous due to neighborhood effect averaging: Implications for
  environmental inequality research.
\newblock {\em Environmental Research}, 195:110519.

\bibitem[Künzli et~al., 2003]{kunzli_breathless_2003}
Künzli, N., McConnell, R., Bates, D., Bastain, T., Hricko, A., Lurmann, F.,
  Avol, E., Gilliland, F., and Peters, J. (2003).
\newblock Breathless in {Los} {Angeles}: {The} {Exhausting} {Search} for
  {Clean} {Air}.
\newblock {\em American Journal of Public Health}, 93(9):1494--1499.

\bibitem[Lee et~al., 2006]{https://doi.org/10.1111/j.1365-2222.2006.02555.x}
Lee, S.~L., Wong, W. H.~S., and Lau, Y.~L. (2006).
\newblock {Association between air pollution and asthma admission among
  children in Hong Kong}.
\newblock {\em Clinical \& Experimental Allergy}, 36(9):1138--1146.

\bibitem[Liévanos, 2019]{lievanos_racialized_2019}
Liévanos, R.~S. (2019).
\newblock Racialized {Structural} {Vulnerability}: {Neighborhood} {Racial}
  {Composition}, {Concentrated} {Disadvantage}, and {Fine} {Particulate}
  {Matter} in {California}.
\newblock {\em International Journal of Environmental Research and Public
  Health}, 16(17):3196.

\bibitem[Lu, 2021]{LU2021111549}
Lu, Y. (2021).
\newblock {Beyond air pollution at home: Assessment of personal exposure to
  PM2.5 using activity-based travel demand model and low-cost air sensor
  network data}.
\newblock {\em Environmental Research}, 201:111549.

\bibitem[Lu et~al., 2021]{LU2021110653}
Lu, Y., Giuliano, G., and Habre, R. (2021).
\newblock {Estimating hourly PM2.5 concentrations at the neighborhood scale
  using a low-cost air sensor network: A Los Angeles case study}.
\newblock {\em Environmental Research}, 195:110653.

\bibitem[Mikati et~al., 2018]{mikati_disparities_2018}
Mikati, I., Benson, A.~F., Luben, T.~J., Sacks, J.~D., and Richmond-Bryant, J.
  (2018).
\newblock Disparities in {Distribution} of {Particulate} {Matter} {Emission}
  {Sources} by {Race} and {Poverty} {Status}.
\newblock {\em American Journal of Public Health}, 108(4):480--485.

\bibitem[Milojevic et~al., 2017]{milojevic2017socioeconomic}
Milojevic, A., Niedzwiedz, C.~L., Pearce, J., Milner, J., MacKenzie, I.~A.,
  Doherty, R.~M., and Wilkinson, P. (2017).
\newblock {Socioeconomic and urban-rural differentials in exposure to air
  pollution and mortality burden in England}.
\newblock {\em Environmental Health}, 16(1):1--10.

\bibitem[Mohl, 2004]{mohl_stop_2004}
Mohl, R.~A. (2004).
\newblock Stop the {Road}: {Freeway} {Revolts} in {American} {Cities}.
\newblock {\em Journal of Urban History}, 30(5):674--706.

\bibitem[Moreno-Jimenez et~al., 2016]{moreno2016assessing}
Moreno-Jimenez, A., Ca{\~n}ada-Torrecilla, R., Vidal-Dom{\'\i}nguez, M.~J.,
  Palacios-Garcia, A., and Martinez-Suarez, P. (2016).
\newblock {Assessing environmental justice through potential exposure to air
  pollution: A socio-spatial analysis in Madrid and Barcelona, Spain}.
\newblock {\em Geoforum}, 69:117--131.

\bibitem[Nyhan et~al., 2016]{doi:10.1021/acs.est.6b02385}
Nyhan, M., Grauwin, S., Britter, R., Misstear, B., McNabola, A., Laden, F.,
  Barrett, S. R.~H., and Ratti, C. (2016).
\newblock {\enquote{Exposure Track}—The Impact of Mobile-Device-Based
  Mobility Patterns on Quantifying Population Exposure to Air Pollution}.
\newblock {\em Environmental Science \& Technology}, 50(17):9671--9681.

\bibitem[Oosterlee et~al., 1996]{oosterlee1996}
Oosterlee, A., Drijver, M., Lebret, E., and Brunekreef, B. (1996).
\newblock Chronic respiratory symptoms in children and adults living along
  streets with high traffic density.
\newblock {\em Occupational and Environmental Medicine}, 53:241--7.

\bibitem[Padilla et~al., 2014]{padilla2014air}
Padilla, C.~M., Kihal-Talantikite, W., Vieira, V.~M., Rossello, P., Le~Nir, G.,
  Zmirou-Navier, D., and Deguen, S. (2014).
\newblock {Air quality and social deprivation in four French metropolitan
  areas: A localized spatio-temporal environmental inequality analysis}.
\newblock {\em Environmental Research}, 134:315--324.

\bibitem[Pan et~al., 2013]{doi:10.1080/10962247.2013.763867}
Pan, H., Bartolome, C., Gutierrez, E., Princevac, M., Edwards, R., Boarnet,
  M.~G., and Houston, D. (2013).
\newblock {Investigation of roadside fine particulate matter concentration
  surrounding major arterials in five Southern Californian cities}.
\newblock {\em Journal of the Air \& Waste Management Association},
  63(4):482--498.

\bibitem[Pastor et~al., 2001]{doi:10.1111/0735-2166.00072}
Pastor, M., Sadd, J., and Hipp, J. (2001).
\newblock {Which Came First? Toxic Facilities, Minority Move-In, and
  Environmental Justice}.
\newblock {\em Journal of Urban Affairs}, 23(1):1--21.

\bibitem[Perez, 2017]{perez_angeles_2017}
Perez, J. (2017).
\newblock The {Los} {Angeles} {Freeway} and the {History} of {Community}
  {Displacement}.
\newblock {\em Toro Historical Review}, 3.

\bibitem[Polichetti et~al., 2009]{polichetti_effects_2009}
Polichetti, G., Cocco, S., Spinali, A., Trimarco, V., and Nunziata, A. (2009).
\newblock Effects of particulate matter ({PM10}, {PM2}.5 and {PM1}) on the
  cardiovascular system.
\newblock {\em Toxicology}, 261(1-2):1--8.

\bibitem[{PolicyLink}, 2017]{policylink_equity_2017}
{PolicyLink} (2017).
\newblock An {Equity} {Profile} of the {Los} {Angeles} {Region}.

\bibitem[Poorfakhraei et~al., 2017]{poorfakhraei_evaluating_2017}
Poorfakhraei, A., Tayarani, M., and Rowangould, G. (2017).
\newblock Evaluating health outcomes from vehicle emissions exposure in the
  long range regional transportation planning process.
\newblock {\em Journal of Transport \& Health}, 6:501--515.

\bibitem[Pulido, 2000]{pulido_rethinking_2000}
Pulido, L. (2000).
\newblock {Rethinking Environmental Racism: White Privilege and Urban
  Development in Southern California}.
\newblock {\em Annals of the Association of American Geographers}, page~30.

\bibitem[Ramezani et~al., 2021]{RAMEZANI202128}
Ramezani, S., Hasanzadeh, K., Rinne, T., Kajosaari, A., and Kyttä, M. (2021).
\newblock Residential relocation and travel behavior change: Investigating the
  effects of changes in the built environment, activity space dispersion, car
  and bike ownership, and travel attitudes.
\newblock {\em Transportation Research Part A: Policy and Practice},
  147:28--48.

\bibitem[Reichmuth, 2019]{reichmuth_inequitable_2019}
Reichmuth, D. (2019).
\newblock {Inequitable Exposure to Air Pollution from Vehicles in California
  Fact Sheet}.
\newblock Technical report, Union of Concerned Scientists.

\bibitem[Rivas et~al., 2017]{rivas2017exposure}
Rivas, I., Kumar, P., and Hagen-Zanker, A. (2017).
\newblock Exposure to air pollutants during commuting in {London}: Are there
  inequalities among different socio-economic groups?
\newblock {\em Environment International}, 101:143--157.

\bibitem[Rowangould et~al., 2016]{rowangould_identifying_2016}
Rowangould, D., Karner, A., and London, J. (2016).
\newblock Identifying environmental justice communities for transportation
  analysis.
\newblock {\em Transportation Research Part A}, 88:151--162.

\bibitem[Rowangould, 2015]{Rowangould2015ANA}
Rowangould, G. (2015).
\newblock A new approach for evaluating regional exposure to particulate matter
  emissions from motor vehicles.
\newblock {\em Transportation Research Part D}, 34:307--317.

\bibitem[Sarzynski, 2012]{sarzynski2012}
Sarzynski, A. (2012).
\newblock {Bigger Is Not Always Better: A Comparative Analysis of Cities and
  their Air Pollution Impact}.
\newblock {\em Urban Studies}, 49(14):3121--3138.

\bibitem[SCAG, 2010]{scag_heavy_2010}
SCAG (2010).
\newblock Heavy {Duty} {Truck} {Model}.
\newblock Technical report, Southern California Association of Governments.

\bibitem[Schindler and Caruso, 2021]{https://doi.org/10.1111/gean.12288}
Schindler, M. and Caruso, G. (2021).
\newblock Urban interventions to reduce pollution exposure and improve spatial
  equity.
\newblock {\em Geographical Analysis}, online before print.

\bibitem[Schweitzer and Valenzuela, 2004]{schweitzer_environmental_2004}
Schweitzer, L. and Valenzuela, A. (2004).
\newblock Environmental {Injustice} and {Transportation}: {The} {Claims} and
  the {Evidence}.
\newblock {\em Journal of Planning Literature}, 18(4):383--398.

\bibitem[Seagram et~al., 2019]{doi:10.1177/0361198119825538}
Seagram, A.~F., Brown, S.~G., Huang, S., Landsberg, K., and Eisinger, D.~S.
  (2019).
\newblock {National Assessment of Near-Road Air Quality in 2016: Multi-Year
  Pollutant Trends and Estimation of Near-Road PM2.5 Increment}.
\newblock {\em Transportation Research Record}, 2673(2):161--171.

\bibitem[Sunyer et~al., 2015]{Sunyer2015}
Sunyer, J., Esnaola, M., Alvarez-Pedrerol, M., Forns, J., Rivas, I.,
  López-Vicente, M., Suades-González, E., Foraster, M., Garcia-Esteban, R.,
  Basagaña, X., and et~al. (2015).
\newblock Association between traffic-related air pollution in schools and
  cognitive development in primary school children: A prospective cohort study.
\newblock {\em PLOS Medicine}, 12(3):e1001792.

\bibitem[Tayarani and Rowangould, 2020]{TAYARANI2020108999}
Tayarani, M. and Rowangould, G. (2020).
\newblock Estimating exposure to fine particulate matter emissions from vehicle
  traffic: Exposure misclassification and daily activity patterns in a large,
  sprawling region.
\newblock {\em Environmental Research}, 182:108999.

\bibitem[Temam et~al., 2017]{temam2017socioeconomic}
Temam, S., Burte, E., Adam, M., Ant{\'o}, J.~M., Basaga{\~n}a, X., Bousquet,
  J., Carsin, A.-E., Galobardes, B., Keidel, D., K{\"u}nzli, N., et~al. (2017).
\newblock {Socioeconomic position and outdoor nitrogen dioxide (NO2) exposure
  in Western Europe: A multi-city analysis}.
\newblock {\em Environment International}, 101:117--124.

\bibitem[Tessum et~al., 2017]{tessum_inmap_2017}
Tessum, C.~W., Hill, J.~D., and Marshall, J.~D. (2017).
\newblock {InMAP}: {A} model for air pollution interventions.
\newblock {\em PLOS ONE}, 12(4):e0176131.

\bibitem[Tessum et~al., 2021]{Tessumeabf4491}
Tessum, C.~W., Paolella, D.~A., Chambliss, S.~E., Apte, J.~S., Hill, J.~D., and
  Marshall, J.~D. (2021).
\newblock {PM2.5 polluters disproportionately and systemically affect people of
  color in the United States}.
\newblock {\em Science Advances}, 7(18).

\bibitem[Thompson, 2018]{thompson_airborne_2018}
Thompson, J.~E. (2018).
\newblock Airborne {Particulate} {Matter}: {Human} {Exposure} and {Health}
  {Effects}.
\newblock {\em Journal of Occupational \& Environmental Medicine},
  60(5):392--423.

\bibitem[{US EPA}, 2021]{us_epa_2021}
{US EPA} (2021).
\newblock Fast {Facts}: {U.S.} {Transportation} {Sector} {Greenhouse} {Gas}
  {Emissions}, 1990-2019.
\newblock Technical report, United States Environmental Protection Agency.

\bibitem[Yuan, 2018]{yuan_mega_2018}
Yuan, Q. (2018).
\newblock Mega freight generators in my backyard: {A} longitudinal study of
  environmental justice in warehousing location.
\newblock {\em Land Use Policy}, 76:130--143.

\bibitem[Zhu et~al., 2002]{PMID:12269664}
Zhu, Y., Hinds, W.~C., Kim, S., and Sioutas, C. (2002).
\newblock Concentration and size distribution of ultrafine particles near a
  major highway.
\newblock {\em Journal of the Air \& Waste Management Association},
  52(9):1032—1042.

\end{thebibliography}

\end{document}